\documentclass[conference]{IEEEtran}
\usepackage{cite}
\usepackage{url}
\usepackage{tikz}
\usepackage{footnote}
\usepackage{balance}

\graphicspath{{./graphics/}}
\DeclareGraphicsExtensions{.pdf,.jpeg,.png}
\usepackage[cmex10]{amsmath}
\usepackage{todonotes}

\usepackage{dblfloatfix}
\usepackage{fixltx2e}

\usepackage{graphicx}
\usepackage{caption}
\usepackage{subcaption}
\usepackage[]{algorithm2e}
\usepackage{amsmath}
\usepackage{amsfonts} 
\usepackage[margin=1.15in]{geometry}
\usepackage{units} 
\usepackage{authblk}

\begin{document}

\title{Graphulo: Linear Algebra Graph Kernels \\ for NoSQL Databases}


\author[*]{Vijay Gadepally\thanks{vijayg@csail.mit.edu}}
\author[$\dagger$]{Jake Bolewski\thanks{jakebolewski@gmail.com}}
\author[*]{Dan Hook\thanks{daniel.hook@ll.mit.edu}}
\author[$\ddagger$]{Dylan Hutchison\thanks{dhutchis@stevens.edu}}
\author[*]{Ben Miller\thanks{bamiller@ll.mit.edu}}
\author[*]{Jeremy Kepner\thanks{kepner@csail.mit.edu}}
\affil[*]{MIT Lincoln Laboratory}
\affil[$\dagger$]{MIT Computer Science and Artificial Intelligence Laboratory}
\affil[$\ddagger$]{Stevens Institute of Technology}


\renewcommand\Authands{, }

%

\maketitle

{\let\thefootnote\relax\footnote{Vijay Gadepally is the corresponding
    author and can be reached at vijayg [at] mit.edu
}}
{\let\thefootnote\relax\footnote{This material is based upon work
    supported by the National Science Foundation under Grant
    No. DMS-1312831. Any opinions, findings, and conclusions or recommendations expressed in this material are those of the author(s) and do not necessarily reflect the views of the National Science Foundation.
}}

\setcounter{footnote}{0}

\newcommand{\numparties}{\ensuremath{\mathcal{P}}}


\begin{abstract}

Big data and the Internet of Things era continue to challenge
computational systems. Several technology solutions such as NoSQL
databases have been developed to deal with this challenge. In order to
generate meaningful results from large datasets, analysts often
use a graph representation which provides an intuitive way to
work with the data. Graph vertices can represent users and
events, and edges can represent the relationship between vertices. Graph algorithms are used to extract meaningful
information from these very large graphs. At MIT, the Graphulo
initiative is an effort to perform graph algorithms
directly in NoSQL databases such as Apache Accumulo or SciDB, which have an
inherently sparse data storage scheme. Sparse matrix operations
have a history of efficient implementations and the Graph Basic Linear
Algebra Subprogram (GraphBLAS) community has developed a set of key
kernels that can be used to develop efficient linear algebra
operations. However, in order to use the GraphBLAS kernels, it is
important that common graph algorithms be recast using the linear
algebra building blocks. In this article, we look at common classes of
graph algorithms and recast them into linear algebra operations using the GraphBLAS building blocks.

\end{abstract}

\IEEEpeerreviewmaketitle


\section{Introduction}
\label{sec:introduction}

The volume, velocity and variety~\cite{laney20013d} of data being collected by today's
systems far outpace the ability to provide meaningful results or
analytics. A common way to represent such large unstructured datasets
is through a graph representation as they provide an intuitive
representation of large data sets. In such a representation, graph vertices can represent users or
events and edges can represent the relationship between
vertices. Many recent efforts have looked at the mapping between
graphs and linear algebra. In such a mapping, graphs are often
represented as sparse arrays such as associative arrays or sparse
matrices using a graph schema. One such effort is the Graph Basic
Linear Algebra Subprogram (GraphBLAS) group which looks to provide a
set of kernels that can be used to cast graph algorithms as sparse linear algebraic
operations \cite{graphblas}. The abilty to represent graph algorithms
as linear algebraic operations can be greatly beneficial for
algorithms scaled for large data volume such as those in ~\cite{bader2014graph, gadepally2014big}. However, for such an
initiative to be successful, it is important that the proposed linear
algebra kernels cover a wide variety of graph algorithms that are
often used by analysts. This article looks at common classes of graph algorithms and
provides an initial set of graph algorithms recast as linear algebraic
operations.

The purpose of our present research effort is to enable graph
algorithms directly on NoSQL (Not Only SQL) databases. Databases such as
Apache Accumulo or SciDB have become a popular alternative to traditional
relational databases due to their high availability, partition
tolerance and performance. NoSQL databases often make use of a key
value store or store information in triples which are similar to the
way sparse matrices are stored~\cite{kepner2014gabb}. We see a large
similarity between our work on performing graph algorithms directly on
NoSQL databases and research on the GraphBLAS specification. The GraphBLAS
community has proposed an initial set of \textit{building blocks}:

\begin{itemize}
\item SpGEMM: Sparse Generalized Matrix Multiply
\item SpM\{Sp\}V: Sparse Matrix (Sparse) Vector Multiply
\item SpEWiseX: Sparse Element-wise Multiplication
\item SpRef: Sparse Reference to a subset
\item SpAsgn: Sparse Assignment to a subset
\item Scale: SpEWiseX with a scalar
\item Apply: Apply a function to each element
\end{itemize}

Further, these kernels have been described to work on alternate semiring
structures such as the tropical semiring which replaces traditional
algebra with the \textit{min} operator and the traditional
multiplication with the \textit{+} operator. This flexibility allows a
wide variety of graph analytics to be represented using the
aforementioned building blocks. Table~\ref{tab:1} summarizes 
classes of graph algorithms that are widely used by the graph
analytics community.

\begin{table*}[t]
  \centering
  \begin{tabular}{| l| p{4cm} | p{5cm} | } \hline
   \textbf{Algorithm Class} & \textbf{Description} & \textbf{Algorithm Examples} \\ \hline
   Exploration \& Traversal  & Algorithms to traverse or search
   vertices & Depth First Search, Breadth First Search \\   \hline     
    
   Subgraph Detection \& Vertex Nomination & Finding subgraphs or
    components within a graph  & K-Truss subgraph detection, Clique detection   \\ \hline
       
   Centrality & Finding important vertices or within a graph  & Betweenness Centrality, Eigen Centrality \\ \hline
   
    Similarity & Finding parts of a graph which are similar in terms
    of vertices or edges & Graph Isomorphism, Jaccard Index, Neighbor
    Matching \\ \hline

    Community Detection & Look for communities (areas of high
    connectedness or similarity) within a graph & Topic Modeling,
    Non-negative matrix factorization (NMF), Principle Component
    Analysis, Singular Value Decomposition \\ \hline

    Prediction & Predicting new or missing edges & Link Prediction,
    Emerging community detection \\ \hline

    Shortest Path & Finding the shortest distance between vertices or
    sets of vertices & Floyd Warshall, Bellman Ford, A* Algorithm,
    Johnson's Algorithm \\ \hline

  \end{tabular}
  \caption{Classes of Graph Algorithms}
  \label{tab:1}
\end{table*}

With the popularity of NoSQL databases and the inherent 
parallels between the representation of data in such databases and
sparse arrays, our research effort looks at determining
how kernels from the GraphBLAS specification can be evaluated on NoSQL
databases. However, in order to ensure that these kernels will be able
to perform common NoSQL database tasks, such as exploration and
community detection, 
it is important that the proposed
kernels are able to express a wide variety of common graph
analytics.

\subsection{The Graphulo Initiative}

Graphulo~\cite{graphuloweb} is an ongoing initiative at the Massachusetts Institute of
Technology that looks at using the GraphBLAS kernels on the Apache
Accumulo database. Accumulo is used for a
variety of applications and has some of the highest published
performance~\cite{kepner2014achieving}. A goal of the Graphulo
initiative is to use Accumulo server components such as iterators to
perform graph analytics. In order to provide end users with a
specification to which they can write their algorithms, Graphulo is
being written to conform to the GraphBLAS specifications.

\subsection{Paper Outline}

In this article, we present an initial set of common graph
algorithms recast in the language of sparse linear algebra and
expressed using the proposed GraphBLAS kernels. In Section~\ref{sec:graphschemas} we introduce
the base datatype of NoSQL databases - associative arrays - and
discuss common schemas used to represent large graphs in
associative arrays. In Section~\ref{sec:algorithms}, we recast
  popular graph algorithms from the Exploration \& Traversal, Subgraph Detection, Centrality and
Community Detection classes of graph algorithms using GraphBLAS
kernels. In Section~\ref{discussion} we discuss the results,
limitations and future work and provide readers with an
understanding of how these algorithms can be implemented on NoSQL
databases such as Apache Accumulo. We conclude the article in Section~\ref{sec:conclusions}.


\section{Associative Arrays and Graph Schemas}
\label{sec:graphschemas}

The Graphulo project looks at how graph algorithms can be performed on
NoSQL databases. Associative arrays are used as the data type for
storing and manipulating a large variety of complex datasets. In order
to represent a dataset using associative arrays, we look at a few
common schemas that can be used.

\subsection{Associative Arrays}

Associative arrays are
used to describe the relationship between multidimensional entities
using numeric/string keys and numeric/string values. Associative
arrays provide a generalization of sparse matrices. Formally, an
associative array \textbf{A} is a map from $d$ sets of keys $K_1 \times K_2 \times ... \times K_d$ to a value set $V$ with a semi-ring structure
$$
{\bf A}: K_1 \times K_2 \times ... \times K_d \rightarrow V,
$$
where $(V,\oplus,\otimes, 0, 1)$ is a semi-ring with addition operator $\oplus$, multiplication operator $\otimes$, additive-identity/multiplicative-annihilator 0, and multiplicative-identity 1.  Furthermore, associative arrays have a finite number of non-zero values which means their support $supp({\bf A})={\bf A}^{-1} (V \backslash \{ 0 \} )$ is finite.

As a data structure, associative arrays returns a value given some
number of keys and constitute a function between a set of tuples and a
value space. In practice, every associative array can be created from
an empty associative array by simply adding and subtracting values. With
this definition, it is assumed that only a finite number of tuples
will have values, and all other tuples will have a default
value of the additive-identity/multiplicative-annihilator 0. Further, the associative array mapping should support
operations that resemble operations on ordinary vectors and matrices
such as matrix multiplication. In practice, associative arrays support a variety of linear algebraic operations
such as summation, union, intersection, and multiplication. Summation of
two associative arrays, for example, that do not have any common row
or column key performs a union of their underlying non-zero keys. 

Graphulo database tables are exactly described using the mathematics of
associative arrays~\cite{kepner2014gabb}. In the D4M schema, a table in
the Accumulo database is an associative array. In
this context, the primary differences between associative arrays and
sparse matrices are: associative array entries always carry their
global row and column labels while sparse matrices do not. Another
difference between associative arrays is that sparse
matrices can have empty rows or columns while associative arrays do
not. For the purposes of this algorithmic work associative arrays are encoded as sparse matrices. 

\subsection{Graph Schemas}

The promise of big data is the ability to correlate diverse
and heterogeneous data sources to reduce the time to insight. 
Correlating this data requires putting data into a common frame of
reference so that similar entities can be
compared. The associative arrays described in the previous subsection
can be used with a variety of NoSQL databases such as Accumulo and
require a schema to convert the dense arbitrary data into a sparse
associative representation. Given the variety of data, there are a few
commonly used graph schemas~\cite{kepner2014gabb} which we discuss below.

\subsubsection{Adjacency Matrix}

In this schema, data is organized as a graph adjacency matrix which
can represent directed or undirected weighted graphs. In this schema,
rows and columns of the adjacency matrix represents vertices, and
values represent weighted edges between vertices. Adjacency matrices
provide a great deal of functionality and are one of the more common
ways to express graphs through matrices. For graph $G=(V,E)$ where
$\bf V=\{v_{1},v_{2},...,v_{n}\}$ and $\bf E=\{e_{1},e_{2},...,e_{m}\}$, the
adjacency matrix $A$ is a $n\times n$ matrix where: \\

$A(i,j) = \left\{\begin{matrix}
\#\mbox{ } edges\mbox{ } from\mbox{ } v_{i}\mbox{ }to\mbox{ } v_{j}, & if\mbox{ } i \neq j \\ 
number\mbox{ }of\mbox{ } self\mbox{ } loops,& if\mbox{ } i = j
\end{matrix}\right.$
\\

\subsubsection{Incidence Matrix}

The incidence matrix representation of a graph can represent
multi-hyper-weighted as well as directed and multi-partite graphs (multiple edges between
vertices, multiple vertices per edge and multiple partitions). The
incidence matrix representation is capable of representing complex
graphs when compared to the adjacency matrix representation. In the
incidence matrix representation, matrix
rows correspond to edges, and matrix columns represent vertices, with
nonzero values in a row indicated vertices associated with the edge.  The value at a particular row-column
pair represents the edge weight and sign is often used to represent direction. There are many representations for
the incidence matrix, and a common format is described below.

For graph $\bf G= (V, E)$ where
$\bf V=\{v_{1},v_{2},...,v_{n}\}$ and $\bf E=\{e_{1},e_{2},...,e_{m}\}$, the
incidence matrix $\bf E$ is a $m\times n$ matrix where: \\

$E(i,j) = \left\{\begin{matrix}
+|e_{i}| & if\mbox{ } e_{i}\mbox{ } goes \mbox{ } into \mbox{ } v_{j}
\\ 
-|e_{i}| & if\mbox{ } e_{i}\mbox{ } leaves \mbox{ } v_{j} \\ 
0 & otherwise
\end{matrix}\right.$
\\

\subsubsection{D4M Schema}

The D4M 2.0 Schema~\cite{kepner2013d4m}, provides a four associative
array solution, ($Tedge$, $Tedge^{\sf T}$, $Tdeg$, and $Traw$), 
that can be used to represent complex data. The edge tables, $Tedge$
and $Tedge^{\sf T}$, contain the full semantic information of the data set
in the rows and columns of the associative arrays. From the schema described in~\cite{kepner2013d4m}, a dense database can be
converted to a sparse representation by exploding each data entry into an associative array
where each unique column-value pair
is a column. The $Tdeg$ array maintains a count of the degrees of each
of the columns of $Tedge$, and $Traw$ is used to store the raw data. A
more thorough description of the schema is provided in~\cite{kepner2013d4m}.
 Once in sparse matrix form, the full machinery of linear
algebraic graph processing and detection theory can be
applied. Linear algebraic operations applied on associative arrays
organized using the D4M schema can have useful results. For example,
addition of two arrays represents a union, and the
multiplication of two arrays represents a correlation.


\section{Algorithms}
\label{sec:algorithms}

There are many different graph algorithms that can be analyzed. In
this section, we present an overview of our work in representing the
classes of graph algorithms presented in Table~\ref{tab:1} using
  kernels from the GraphBLAS specification. For the work presented in
  this section, we encode associative arrays as sparse matrices.

\subsection{Centrality}
Of the many centrality metrics, there are a few that are particularly
well-suited to the GraphBLAS framework. Degree centrality, for
example, assumes that a vertex's importance is proportional to the
number of connections it shares. Given an adjacency matrix, $\bf A$, this can easily be computed via a row or column reduction, depending on whether in- or out-degree is of interest.

Other centrality metrics are explicitly linear algebraic in their
formulation. For example, eigenvector centrality assumes that each
vertex's centrality is proportional to the sum of its neighbors'
centrality scores. This is equivalent to scoring each vertex based on
its corresponding entry in the principal eigenvector, which can be
computed via the power method. Starting with Starting with a random positive vector $x_0$ with entries between zero and 1,
we iteratively compute $$x_{k+1} = \mathbf{A}x_k$$ until $|x_{k+1}^Tx_k|/(\|x_{k+1}\|_2\|x_k\|_2)$ is close to 1. 

Related metrics are Katz centrality and PageRank. Katz centrality considers the number of $k$-hop paths to a vertex, for all $k$, penalizing those with higher distances. This is also computed via an iterative procedure in which the $k$th-order degree vector is computed, and added to an accumulator as follows:
\begin{align}
d_{k+1}& = \mathbf{A} d_k \nonumber\\ 
x_{k+1}& = x_k + \alpha^kd_{k+1}, \nonumber
\end{align}
where $d_0$ is a vector of $1$s and  we use the same stopping
criterion as eigenvector centrality. PageRank simulates a random walk
on a graph, with the possibility of jumping to an arbitrary
vertex. Each vertex is then ranked according to the probability of
landing on it at an arbitrary point in an infinite random walk. If the
probability of jumping to an arbitrary vertex is 0, then this is
simply the principal eigenvector of $\bf A^{\sf T}  \bf D^{-1}$, where $\bf
D$ is a diagonal matrix of vertex out-degrees. If the probability of a
jump is $\alpha$, then we compute the principal eigenvector
of $$\frac{\alpha}{N} \mathbf{1}_{N\times N}+(1-\alpha) \bf A^{\sf T}
\bf D^{-1}.$$ As with eigenvector centrality, this can be done using
the power method, where multiplication by a matrix of $\bf 1$s can be emulated by summing the vector entries and creating a new vector where each entry is equal to the resulting value. All of these centrality measures rely on doing iterative matrix-vector multiplications, which fits nicely within the scope of GraphBLAS.

There has also been work on casting betweenness centrality---where a vertex's importance is based on the number of shortest paths that contain it---in linear-algebraic operations~\cite{kepner2011graph}. Other metrics, such as closeness centrality, will be the subject of future work.

\subsection{Subgraph detection and vertex nomination}
Detection of interesting and anomalous subgraphs has been a problem of interest for the computer science community for many years. Examples of this problem space include vertex nomination (ranking vertices based on how likely they are to be associated with a subset of ``cue'' vertices) \cite{Coppersmith12}, planted clique detection \cite{NadakuditiPlantedClique}, and planted cluster detection \cite{AriasCastro13}.

A problem related to planted clique and planted cluster detection is computing the truss decomposition. A $k$-truss is a graph in which every edge is part of at least $k-2$ triangles. Any graph is a 2-truss, and any $k$-truss in a graph is part of a $(k-1)$-truss in the same graph. Computing the truss decomposition of a graph involves finding the maximal $k$-truss for all $k\geq2$. A recent technique for computing the truss decomposition \cite{WangChenTruss} can be easily converted into linear-algebraic operations. Define the support of an edge to be the number of triangles of which the edge is a member. The algorithm can be summarized as follows:
\begin{enumerate}
\item Compute the support for every edge.
\item If there is no edge with support less than $k-2$, stop\label{check}.
\item Otherwise, remove an edge with support less than $k-2$, update the supports of its associated vertices, and go to \ref{check}.
\end{enumerate}
In \cite{WangChenTruss}, a more efficient algorithm is proposed that considers the edges in order of increasing support. In the linear-algebraic form, all edges are considered at once, and the appropriate edges are removed simultaneously.

To see the linear-algebraic algorithm, first consider the unoriented
incidence matrix $\bf E$. Each row of $\bf E$ has a $1$ in the
column of each associated vertex. To get the support for this edge, we
need the overlap of the neighborhoods of these vertices. If the rows
of the adjacency matrix \textbf{A} associated with the two vertices
are summed, this corresponds to the entries that are equal to
2. Summing these rows is equivalent to multiplying \textbf{A} on the
left by the edge's row in $\bf E$. Therefore, to get the support for each
edge, we can compute $\bf EA$, apply to each entry a function that maps 2
to 1 and all other values to 0, and sum each row of the resulting
matrix. Note also that $$\bf A= \mathbf{E^{\sf T}} \mathbf{E}- diag(\mathbf{E^{\sf T}E}),$$ which allows us to
recompute $\bf EA$ after edge removal without performing the full matrix
multiplication. We take advantage of this fact in
Algorithm~\ref{ktruss}. Within the pseudocode, $x_c$ refers to the
complement of $x$ in the set of row indices. This algorithm can return
the full truss decomposition by computing the truss with $k=3$ on the
full graph, then passing the resulting incidence matrix to the
algorithm with an incremented $k$. This procedure will continue until
the resulting incidence matrix is empty. This algorithm can be realized
using the GraphBLAS kernels SpGEMM, SpMV, and Apply. 

\begin{algorithm}
 \KwData{The unoriented incidence matrix $\bf E$, integer $k$}
 \KwResult{Incidence matrix of k-truss subgraph $E_{k}$}
 initialization\;
	$d = \operatorname{sum}(E)$\\
        $A =  E^{\sf T}E - diag(d)$\\
        $R = EA$ \\
	$s = (R ==2)\mathbf{1}$\\
        $x = find(s < k-2)$ \\
 \While{x is not empty}{
  $ E_{x} = E(x,:)$ \\
  $ E = E(x_{c},:)$\\
  $d_x = \operatorname{sum}(E_x)$\\
  $R=R(x_{c},:)$ \\
  $R=R-E [E_{x}^{\sf T}E_{x} - diag(d_x)]$ \\
  $s = (R==2)\mathbf{1}$\\
  $x = find(s < k-2)$ \\
 }
\Return $\bf E$
\vspace{10pt}
 \caption{Algorithm to compute $k$-truss using linear
   algebra. \textbf{1} refers to an array of $1$s}
\label{ktruss}
\end{algorithm}

\begin{figure}
\begin{centering}
\includegraphics[width=2in]{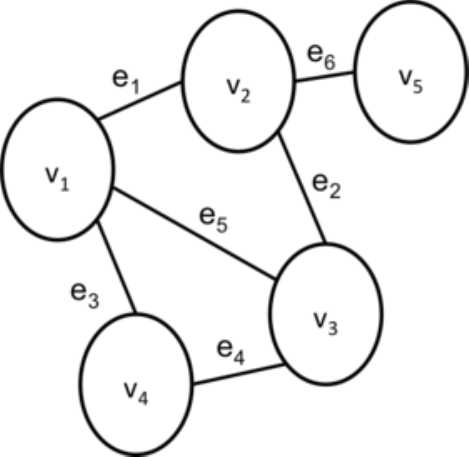}
\caption{Example 5-vertex graph}
\label{examplegraph}
\end{centering}
\end{figure}

As an example of computing the $k$-truss using the algorithm described,
consider the task of finding the 3-truss of the graph in
Fig.~\ref{examplegraph}. 

The incidence matrix for the graph shown in Figure~\ref{examplegraph} is
\begin{eqnarray}
\bf E= \begin{bmatrix}
1 & 1 & 0 & 0 & 0\\ 
0 & 1 & 1 & 0 & 0\\ 
1 & 0 & 0 & 1 & 0\\ 
0 & 0 & 1 & 1 & 0\\ 
1 & 0 & 1 & 0 & 0\\ 
0 & 1 & 0 & 0 & 1 
\end{bmatrix}. \nonumber
\end{eqnarray}
From $\bf E$, we can compute \textbf{A} using the relation ${A = E^{T}E - diag(d)}$ to be:
\begin{eqnarray}
\bf A = \begin{bmatrix}
3 & 1 & 1 & 1 & 0\\ 
1 & 3 & 1 & 0 & 1\\ 
1 & 1 & 3 & 1 & 0\\ 
1 & 0 & 1 & 2 & 0\\ 
0 & 1 & 0 & 0 & 1\\ 
\end{bmatrix}-\begin{bmatrix}
3 & 0 & 0 & 0 & 0\\ 
0 & 3 & 0 & 0 & 0\\ 
0 & 0 & 3 & 0 & 0\\ 
0 & 0 & 0 & 2 & 0\\ 
0 & 0 & 0 & 0 & 1\\ 
\end{bmatrix}. \nonumber
\end{eqnarray}
To get the support, we first compute
\begin{eqnarray}
R &=&\begin{bmatrix}
1 & 1 & 0 & 0 & 0\\ 
0 & 1 & 1 & 0 & 0\\ 
1 & 0 & 0 & 1 & 0\\ 
0 & 0 & 1 & 1 & 0\\ 
1 & 0 & 1 & 0 & 0\\ 
0 & 1 & 0 & 0 & 1 
\end{bmatrix}\begin{bmatrix}
0 & 1 & 1 & 1 & 0\\ 
1 & 0 & 1 & 0 & 1\\ 
1 & 1 & 0 & 1 & 0\\ 
1 & 0 & 1 & 0 & 0\\ 
0 & 1 & 0 & 0 & 0\\ 
\end{bmatrix}\nonumber\\
&=&\begin{bmatrix}
1 & 1 & 2 & 1 & 1\\ 
2 & 1 & 1 & 1 & 1\\ 
1 & 1 & 2 & 1 & 0\\ 
2 & 1 & 1 & 1 & 0\\ 
1 & 2 & 1 & 2 & 0\\ 
1 & 1 & 1 & 0 & 1 
\end{bmatrix}. \nonumber
\end{eqnarray}
The support is then given by
\begin{eqnarray}
s = (R==2)\mathbf{1} =\begin{bmatrix}
0 & 0 & 1 & 0 & 0\\ 
1 & 0 & 0 & 0 & 0\\ 
0 & 0 & 1 & 0 & 0\\ 
1 & 0 & 0 & 0 & 0\\ 
0 & 1 & 0 & 1 & 0\\ 
0 & 0 & 0 & 0 & 0 
\end{bmatrix}\begin{bmatrix}
1 \\
1 \\
1 \\
1 \\
1 \\
\end{bmatrix}=\begin{bmatrix}
1 \\
1 \\
1 \\
2 \\
0 \\ 
\end{bmatrix}. \nonumber
\end{eqnarray}
Since $k=3$, $x$ will be the set of edges where the support is less
than 1, i.e., $x=\{6\}$ and $x_c=\{1,2,3,4,5\}$. Thus, $R$ and $E$ will be set to their first $5$ rows, and the update will be computed as follows:
\begin{eqnarray}
R &=& \begin{bmatrix}
1 & 1 & 2 & 1 & 1\\ 
2 & 1 & 1 & 1 & 1\\ 
1 & 1 & 2 & 1 & 0\\ 
2 & 1 & 1 & 1 & 0\\ 
1 & 2 & 1 & 2 & 0
\end{bmatrix} \nonumber \\
 &\ &-\begin{bmatrix}
1 & 1 & 0 & 0 & 0\\ 
0 & 1 & 1 & 0 & 0\\ 
1 & 0 & 0 & 1 & 0\\ 
0 & 0 & 1 & 1 & 0\\ 
1 & 0 & 1 & 0 & 0
\end{bmatrix}\begin{bmatrix}
0 & 0 & 0 & 0 & 0\\ 
0 & 0 & 0 & 0 & 1\\ 
0 & 0 & 0 & 0 & 0\\ 
0 & 0 & 0 & 0 & 0\\ 
0 & 1 & 0 & 0 & 0\\ 
\end{bmatrix}\nonumber\\
&=&\begin{bmatrix}
1 & 1 & 2 & 1 & 0\\ 
2 & 1 & 1 & 1 & 0\\ 
1 & 1 & 2 & 1 & 0\\ 
2 & 1 & 1 & 1 & 0\\ 
1 & 2 & 1 & 2 & 0
\end{bmatrix}. \nonumber
\end{eqnarray}
The pattern of $2$s in $R$ did not change with the removal of edge 6, so the support will not change. Therefore, the graph represented by the new incidence matrix is a $3$-truss.

\subsection{Similarity}
Computing vertex similarity is important in applications such as link prediction \cite{LinkPrediction07}. One common method for determining the similarity of two vertices is to compute the Jaccard coefficient. This quantity measures the overlap of the neighborhoods of two vertices in an unweighted, undirected graph. 
For vertices $v_{i}$ and $v_{j}$ where $N(v)$ denotes the neighbors of vertex $v$, the Jaccard coefficient is defined as
\begin{eqnarray}
J_{ij} =\frac{\mid N(v_{i}) \bigcap N(v_{j}) \mid}{\mid N(v_{i}) \bigcup N(v_{j}) \mid}.
\label{jaccardeqn}
\end{eqnarray}
Given the connection vectors (a column or row in the adjacency matrix \textbf{A}) for vertices $v_{i}$ and $v_{j}$ (denoted as $a_{i}$ and $a_{j}$) the numerator and denominator of Equation~\ref{jaccardeqn} can be expressed as $a_{i}^{\sf T}a_{j}$ where we replace multiplication with the AND operator in the numerator and the OR operator in the denominator. This gives us
\begin{eqnarray}
J_{ij} &=&(a_{i}^{\sf T}  \wedge a_{j}) ./ (a_{i}^{\sf T}  \vee a_{j}) \nonumber \\
J_{ij} &=&\bf A^{2}_{AND} ./ \bf A^{2}_{OR}. \nonumber 
\end{eqnarray}
This, however, would involve computing a dense matrix, and we are
primarily interested in cases where this is impossible. Two phenomena
can be exploited that will help provide an efficient implementation:
the symmetry of $J$ and sparseness of $\bf A_{AND}^2$. Since $J$ is
symmetric, we can compute only the upper triangular part and then add
the transpose. First we compute the upper triangular part of the
numerator in the entry wise division. The numerator is $\bf A_{AND}^2$, which in an unweighted graph is the same as computing a standard matrix multiplication. We can represent \textbf{A} as $L+U$, where $L$ is strictly lower triangular and $U$ is strictly upper triangular. Since \textbf{A} is symmetric, $L=U^{\sf T}$. Thus, we have 
\begin{eqnarray}
\mathbf{A}^2=(L+U)^2&=&L^2+LU+UL+U^2 \nonumber \\
&=&(U^2)^{\sf T} +U^2+U^{\sf T} U+UU^{\sf T} \nonumber
\end{eqnarray}
It can be verified that $U^2$ is strictly upper triangular and,
therefore $(U^2)^{\sf T}$ is strictly lower triangular. After we compute the
upper triangular part of $\bf A^2$, we can divide each nonzero value by
the number of total neighbors of the associated vertices. Exploiting
these properties, we can compute the Jaccard coefficient as described
in Algorithm \ref{jaccard}. The triu operation extracts the upper
triangular part of the graph, as in MATLAB. Algorithm ~\ref{jaccard}
    can be computed using the GraphBLAS kernels SpGEMM,
    SpMV, and SpEWiseX. Computing the upper triangular part of a graph can be done
    through a user-defined function that implements the Hadamard
    product. For example, if $\otimes=f(i,j)$, $triu(\bf A)=\bf A \otimes 1$
    where $ f(i,j) = \{A(i,j): i \leq j, 0 \mbox{ } otherwise \} $. An example of the computation on the graph in Fig. \ref{examplegraph} is provided in Fig. \ref{examplejaccard}.

\begin{algorithm}
 \KwData{Adjacency matrix $\bf A$}
 \KwResult{Matrix of Jaccard indices $J$}
 initialization\;
	$d = \operatorname{sum}(\bf A)$\\
        $U = \operatorname{triu}(\bf A)$\\
	$X = UU^T$\\
	$Y = U^TU$\\
	$J = U^2+\operatorname{triu}(X)+\operatorname{triu}(Y)$\\
        $J = J - diag(J)$\\
 \For{each nonzero entry $J_{ij}$ in $J$}{
  $J_{ij}=J_{ij}/(d_{i} + d_{j} - J_{ij})$ 
 }
 $J = J + J^{\sf T}$ 
\vspace{10pt}

 \caption{Algorithm to compute Jaccard index using linear algebra.}
\label{jaccard}
\end{algorithm}

\begin{figure*}
\begin{align*}
U &=
\left[
\begin{array}{ccccc}
 0 & 1 & 1 & 1 & 0 \\
 0 & 0 & 1 & 0 & 1 \\
 0 & 0 & 0 & 1 & 0 \\
 0 & 0 & 0 & 0 & 0 \\
 0 & 0 & 0 & 0 & 0 \\
\end{array}
\right] \qquad
U^2 =
\left[
\begin{array}{ccccc}
 0 & 0 & 1 & 1 & 1 \\
 0 & 0 & 0 & 1 & 0 \\
 0 & 0 & 0 & 0 & 0 \\
 0 & 0 & 0 & 0 & 0 \\
 0 & 0 & 0 & 0 & 0 \\
\end{array}
\right] \qquad
UU^T =
\left[
\begin{array}{ccccc}
 3 & 1 & 1 & 0 & 0 \\
 1 & 2 & 0 & 0 & 0 \\
 1 & 0 & 1 & 0 & 0 \\
 0 & 0 & 0 & 0 & 0 \\
 0 & 0 & 0 & 0 & 0 \\
\end{array}
\right] \\
U^TU &=
\left[
\begin{array}{ccccc}
 0 & 0 & 0 & 0 & 0 \\
 0 & 1 & 1 & 1 & 0 \\
 0 & 1 & 2 & 1 & 1 \\
 0 & 1 & 1 & 2 & 0 \\
 0 & 0 & 1 & 0 & 1 \\
\end{array}
\right] \qquad
J\;\;= 
\left[
\begin{array}{ccccc}
 0 & 1 & 2 & 1 & 1 \\
 0 & 0 & 1 & 2 & 0 \\
 0 & 0 & 0 & 1 & 1 \\
 0 & 0 & 0 & 0 & 0 \\
 0 & 0 & 0 & 0 & 0 \\
\end{array}
\right] \\
J &= 
\left[
\begin{array}{ccccc}
 0 & 1 & 2 & 1 & 1 \\
 0 & 0 & 1 & 2 & 0 \\
 0 & 0 & 0 & 1 & 1 \\
 0 & 0 & 0 & 0 & 0 \\
 0 & 0 & 0 & 0 & 0 \\
\end{array}
\right] 
./
\left(
\left[
\begin{array}{ccccc}
 0 & 3 & 3 & 3 & 3 \\
 0 & 0 & 3 & 3 & 3 \\
 0 & 0 & 0 & 3 & 3 \\
 0 & 0 & 0 & 0 & 2 \\
 0 & 0 & 0 & 0 & 0 \\
\end{array}
\right]
+
\left[
\begin{array}{ccccc}
 0 & 3 & 3 & 2 & 1 \\
 0 & 0 & 3 & 2 & 1 \\
 0 & 0 & 0 & 2 & 1 \\
 0 & 0 & 0 & 0 & 1 \\
 0 & 0 & 0 & 0 & 0 \\
\end{array}
\right]
-
\left[
\begin{array}{ccccc}
 0 & 1 & 2 & 1 & 1 \\
 0 & 0 & 1 & 2 & 0 \\
 0 & 0 & 0 & 1 & 1 \\
 0 & 0 & 0 & 0 & 0 \\
 0 & 0 & 0 & 0 & 0 \\
\end{array}
\right] 
\right)\\
&=
\left[
\begin{array}{ccccc}
 0 & \nicefrac{1}{5} & \nicefrac{1}{2} & \nicefrac{1}{4} & \nicefrac{1}{3} \\
 0 & 0 & \nicefrac{1}{5} & \nicefrac{2}{3} & 0 \\
 0 & 0 & 0 & \nicefrac{1}{4} & \nicefrac{1}{3} \\
 0 & 0 & 0 & 0 & 0 \\
 0 & 0 & 0 & 0 & 0 \\
\end{array}
\right] 
\end{align*}
\caption{Computing Jaccard coefficients of the graph in
Fig. \ref{examplegraph}. 
In line 2, $J= U^2 + \operatorname{triu}{(UU^T)} + \operatorname{triu}{(U^TU)}$. 
In line 3, $J=J./(d_i+d_j-J)$. Computing  $J = J + J^{\sf T}$ removes the order dependence.
Computation is on non-zero entries in each matrix.}
\label{examplejaccard}
\end{figure*}

\subsection{Community Detection}

Community detection is a class of graph algorithms designed to find
community structures within a graph. Graph communities often contain
dense internal connections and may possibly overlap with other
communities. Real graphs such as social media have been shown to
exhibit such community structure on geography, language, age group, etc. \cite{fortunato2010community}. The communities may then be used
to suggest or recommend new information, connections, or even products
as recommender systems do for popular online marketplaces such as
Amazon and Google \cite{schafer1999recommender}. One common method
used as a basis for such systems is topic modeling. Topic modeling is
a very popular class of algorithms that provides an intuitive
look into the topics that make up data.  As an example, consider a set of documents
made up of various terms. Application of topic modeling can automatically determine a set of
topics, the terms that make up a topic and the documents that strongly
align with these topics. Techniques such as topic modeling have gained
wide usage for automatic summarization, document modeling and can
provide users with simple and quick insight into a dataset. Topic
modeling is a general field, and a popular technique for topic modeling
is non-negative matrix factorization \cite{wang2011community,kim2008toward}.

Non-negative matrix factorization (NMF) is a class of tools used to
factorize a given matrix into two matrices. Multiplying these two
matrices produces an approximation of the original matrix. Consider a
matrix \textbf{$A_{m \times n}$} to
be factored into matrices $W_{m \times k}$ and $H_{k \times n}$ where $m$
corresponds to the number of rows of $A$, $n$ corresponds to the
number of columns in \textbf{A}, and $k$ corresponds to the number of
topics. Further, NMF enforces the constraint that none of these
matrices contain any negative elements. 

By definition,

\begin{equation}
\mathbf{A} = \mathbf{W}*\mathbf{H}.
\end{equation}

\noindent In the above factorization, the columns of $\bf W$ can be considered a basis
for the matrix \textbf{A} with the rows of $\bf H$ being the associated weights
needed to reconstruct \textbf{A}. The property that $\bf W$ and $\bf H$ are
nonnegative is useful because it can have physical significance (as
opposed to negative weights or basis elements). One way to find the
matrices $\bf W, H$ such that $\mathbf{A} \approx \mathbf{W}*\mathbf{H}$ is through an iterative
technique such as the algorithm presented in Algorithm~\ref{nmf}.

\begin{algorithm}
 \KwData{Incidence Matrix \textbf{A} (size $m \times n$), number of topics $k$}
 \KwResult{$\bf W$and $\bf H$}
 initialization\;
W = random m x k matrix\\
 \While{$\| A - W*H \|_{F} >  \epsilon$}{
  Solve $W^{\sf T} *W*H$ =$W^{\sf T} *A$ for $H$ \\
  Set elements in $H < 0$ to 0 \\
  Solve $H*H^{\sf T}*W^{\sf T} $ = $H*A^{\sf T} $ for $W$ \\
  Set elements in $W < 0$ to 0 \\
 }
\vspace{10pt}

 \caption{NMF through Iteration. At each step of the iteration, we
   check if the Frobenius norm of the difference between \textbf{A}
   and $\bf W*H$ is
   less than the acceptable error.}
\label{nmf}
\end{algorithm}

In order to solve the equations in Algorithm~\ref{nmf}, it is
necessary to find a least squares solution to a system of linear
equations for $\bf W$ and $\bf H$. One way of doing this is by finding 
the matrix inverse of $W^{\sf T}*W$ and  $H*H^{\sf T}$ (both are square matrices) and multiplying with
the right hand side of the equations. One method to find the matrix inverse is
typically done by techniques such as the Singular Value Decomposition
(SVD). However, in order to make use of the GraphBLAS kernels, we
present an technique used by iterative eigenvalue solvers. In such
systems, for iteration $k$: $X_{k+1} =X_{k}*(2I - AX_{k})$. The
algorithm used to find the matrix inverse for a square matrix $\bf A$ is
given in Algorithm~\ref{inv}.

\begin{algorithm}
 \KwData{Matrix \textbf{A} to invert}
 \KwResult{$X = A ^{-1}$}
 initialization\;
  $\|A_{row} \| = \max_{i} (\sum_{j} A_{ij})$ \\
  $\|A_{col} \| = \max_{j} (\sum_{i} A_{ij})$ \\
 $X_{1}$ = $ A^{\sf T} / ( \|A_{row} \| * \| A_{col} \| ) $ \\
 \While{$\|X_{t+1} - X_{t} \|_{F} > \epsilon$}{
   $X_{t+1}= X_{t}*(2*I_{n \times n} - A*X_{t}) $
 }
\vspace{10pt}

 \caption{Matrix inverse through Iteration. At each iteration, we
   check if the value of $X_{t+1}$ is close to the previous iteration
   estimate of $X$.}
\label{inv}
\end{algorithm}

Using this formulation, computing the
inverse of a matrix can be done purely using GraphBLAS
kernels. Combining Algorithms~\ref{nmf} and \ref{inv}, we can find
compute the NMF of a matrix
\textbf{A} using only GraphBLAS kernels. Where $(W^{\sf T}*W)^{-1}$ and $(H*H^{\sf T})^{-1}$ are determined by
using the relation develop in Algorithm~\ref{inv}.

\begin{algorithm}
 \KwData{Incidence Matrix \textbf{A} (size $m \times n$), number of topics $k$}
 \KwResult{$\bf W$ and $\bf H$}
W = random m x k matrix\\
 \While{$\| A-W*H \|_{F} >  \epsilon$}{
  Solve $H$ =$(W^{\sf T}*W)^{-1}*W^{\sf T}*A$ for $H$ \\
  Set elements in $H < 0$ to 0 \\
  Solve $W^{T}$ = $(H*H^{\sf T})^{-1}*H*A^{\sf T}$ for $W$ \\
  Set elements in $W < 0$ to 0 \\
 }
\vspace{10pt}
 \caption{NMF and Inverse through Iteration.}
\label{nmfinv}
\end{algorithm}

In fact, computing the NMF of a matrix using Algorithm~\ref{nmfinv}
will require the GraphBLAS SpRef/SpAsgn, SpGEMM, Scale, SpEWiseX, and
Reduce kernels. The outlined algorithm has been tested against a social
media dataset and provides intuitive results.

\begin{figure*}[ht!]
\centerline{
\includegraphics[width=6in]{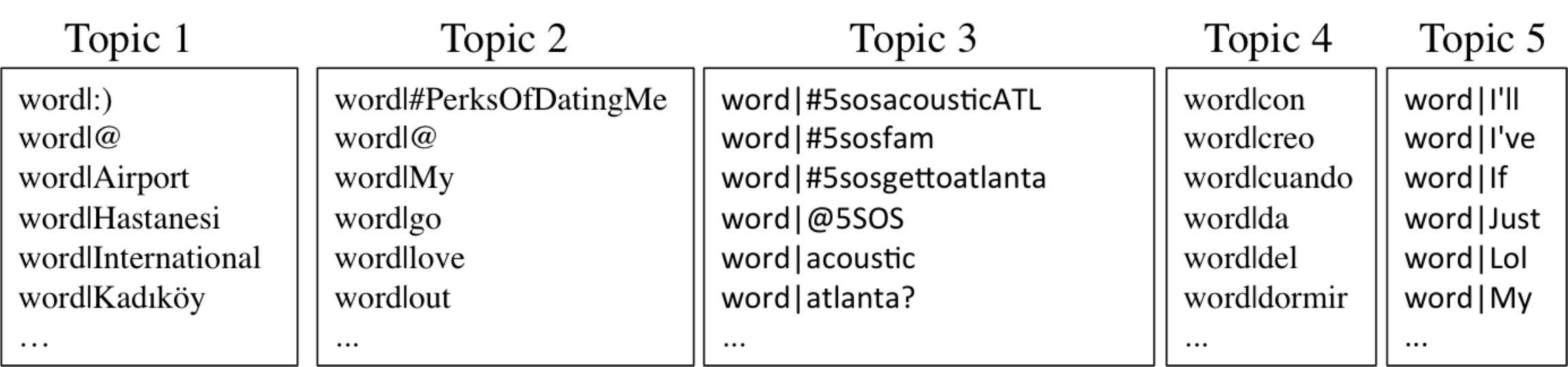}
}
\caption{Application of algorithm~\ref{nmfinv} to 20k tweets and
  modeling with 5 topics. Topic 1 represents tweets with
  Turkish words; topic 2 represents tweets related to dating; topic 3
  relates to an acoustic guitar competition in Atlanta, GA; topic 4
  relates to tweets with Spanish words; and topic 5 represents tweets
  with English words.}
\label{nmf_tweet}
\vspace{-5pt}
\end{figure*}

For example, Algorithm~\ref{nmfinv} was applied to a set of words
collected from the popular social media website Twitter. The algorithm
was used to determine common themes from approximately 20,000 tweets. By setting the number of topics to $5$, we were able to
determine words/tweets that fell into $5$ different topics. The results
from this experiment are shown in Fig.~\ref{nmf_tweet}. From a graph
perspective, this implies that tweets corresponding to these tweets
from a community. For topic 1, as an example, this community represents users who
tweet in the Turkish language.






\section{Discussion}
\label{discussion}
The algorithms presented in this paper demonstrate several algorithmic capabilities using the initial set of GraphBLAS operations, but there are a few inefficiencies that could be improved upon with some additional functions. In Algorithm \ref{ktruss}, for example, when $EA$ is computed, it would be more efficient to only consider the additions that yield a $2$ in the resulting matrix. This could be achieved by replacing the + operator with a logical AND, but this would violate the semiring axioms. Enabling the ability to use linear-algebraic machinery with data operations that do not conform to the rules for semirings may provide substantial speedups.

Algorithm \ref{jaccard} leverages the symmetry of the graph to save some of the unnecessary operations, but some values under the main diagonal must still be computed in the process. Since it is fairly common to work with undirected graphs, providing a version of matrix multiplication that exploits the symmetry, only stores the upper-triangular part, and only computes the upper-triangular part of pairwise statistics, would be a welcome contribution to this effort.

Algorithm \ref{nmfinv} computes the NMF of a matrix $A$ which can
represent the adjacency matrix of a graph. However, calculation of the
matrix inverse using this method can result in dense matrix
operations. Since the aim of this step is to solve a least squares
problem, it would be more efficient to implement this using a sparse
QR factorization or iterative method that preserves the sparsity of the problem as much as
possible. We would welcome community involvement in building these
methods using the GraphBLAS kernels.

As a next step in the Graphulo effort, we will extend the sparse
matrix implementations of the algorithms discussed in this article to
associative arrays. The ability to perform the graph algorithms
described directly on associative arrays will allow us to implement efficient
GraphBLAS operations directly on Accumulo data structures. In order to make efficient implementations, we will use various Accumulo features, 
such as the Accumulo iterator framework, to quickly scan Accumulo tables 
over servers in parallel and perform batch operations such as scaling.

\section{Conclusions}
\label{sec:conclusions}

There are a large variety of graph algorithms that can be used to
solve a diverse set of problems. The Graphulo initiative at the
Massachusetts Institute of Technology is interested in applying the
sparse linear algebra kernels of the GraphBLAS specification to
associative arrays which exactly describe NoSQL database tables such
as those found in the open source Apache Accumulo. Current ongoing
work includes defining efficient implementations of the algorithms
discussed in this article,
extending the classes of supported algorithms and providing a library
that can perform basic operations directly in NoSQL databases.

\section*{Acknowledgment}

The authors wish to thank the Graphulo team at MIT CSAIL and
Lincoln Laboratory. We also thank the reviewers, GraphBLAS contributors and
National Science Foundation for their generous ongoing support of this program.



%

%
%

\bibliographystyle{IEEEtran}
\balance

{\footnotesize
\bibliography{10_bibliography}}

\end{document}